\documentclass{sig-alternate}

\usepackage{colortbl}
\usepackage{cite}
\usepackage{url}
\usepackage{balance}

\definecolor{Gray}{gray}{0.85}

\usepackage{xspace}

\newcommand{\minisec}[1]{\vspace{0.5em}\noindent \textbf{#1}\quad}

\newtheorem{question}{RQ}

\newcommand{\ie}{i.\,e.,\ }
\newcommand{\eg}{e.\,g,\ }
\newcommand{\cf}{c.\,f.,\ }

\begin{document}
%
\conferenceinfo{ICSE}{'10, May 2-8 2010, Cape Town, South Africa}
\CopyrightYear{2010}
\crdata{978-1-60558-719-6/10/05}

\title{Can Clone Detection Support Quality Assessments \\
  of Requirements Specifications?%
\titlenote{This work has
partially been supported by the German Federal Ministry of Education and
Research (BMBF) in the project Quamoco (01 IS 08023F and 01 IS 08023B).}
}

\numberofauthors{2}

\author{
  \alignauthor Elmar Juergens, Florian Deissenboeck,\\ Martin Feilkas, Benjamin Hummel,\\ Bernhard Schaetz, Stefan Wagner\\
    \affaddr{Technische Universit\"at M\"unchen}\\
    \affaddr{Garching b.\ M\"unchen, Germany}
      \and
  \alignauthor Christoph Domann, Jonathan Streit\\
    \affaddr{itestra GmbH}\\
    \affaddr{Garching b. M\"unchen, Germany}
}

\maketitle

\begin{abstract}

Due to their pivotal role in software engineering, considerable
effort is spent on the quality assurance of software requirements
specifications. As they are mainly described in
natural language, relatively few means of automated quality assessment
exist. However, we found that \emph{clone detection}, a technique widely
applied to source code, is promising to  assess one important
quality aspect in an automated way, namely redundancy that stems from
copy\&paste operations. This paper describes a large-scale case study that applied clone detection
to 28 requirements specifications with a total of 8,667 pages. We report on
the amount of redundancy found in real-world specifications, discuss its
nature as well as its consequences and evaluate in how far existing code
clone detection approaches can be applied to assess the quality of
requirements specifications in practice.

\end{abstract}

\category{D.2.1}{Software Engineering}{Requirements/Spe\-ci\-fi\-ca\-tions}
\category{D.2.8}{Software Engineering}{Metrics}[Pro\-duct metrics]

\terms{Documentation, Experimentation, Measurement}

\keywords{Redundancy, Requirements Specification, Clone Detection}

\section{Introduction\label{sec:intro}}

Software requirements specifications (SRS) are the cornerstone of most
software development projects. Due to their pivotal role in the development
process, they have a strong impact on the quality of the developed product
and on the effort spent on
development~\cite{2001_boehmb_top_10,1981_glassr_re_defects}. Moreover,
they are usually the central, and often sole, communication artifact used
between customer and contractor. Thus, the quality of SRS is of paramount
importance for the success of software development projects. However, SRS
are mostly written in natural language and, hence, relatively few
techniques for automated quality assessment exist. Due to the size of
real-world SRS, which often consist of several hundred pages, this poses a
major problem in practice.

One quality defect, \emph{redundancy}, however stands out as it may be
tackled using \emph{clone detection}, a technique that is commonly applied
to find duplications in source code \emph{(cloning)}.  Duplication in
source code has been recognized as a problem for software maintenance by
both the research community and
practitioners~\cite{Koschke2008_clone_identification_removal} as it
increases program size and thus the effort for size-related activities
such as inspections. Moreover, it increases the effort required for
modification, since a change performed on a piece of code often needs to
be performed on its duplicates as well. Furthermore, unintentionally
inconsistent changes to cloned code frequently lead to
errors~\cite{juergens09}. Most of the negative effects of cloning in
programs also hold for cloning in requirements specifications. As SRS are
read and changed often (\eg for requirements elicitation, software design,
and test case specification), redundancy is considered an obstacle to
requirements modifiability~\cite{ieee98} and listed, for instance, as a
major problem in automotive requirements engineering~\cite{WW02}.

\quad Nevertheless, cloning has not been thoroughly investigated in the
context of requirements specifications yet. To remedy this situation, this
paper presents the results of a large-scale case study. The study was
undertaken to find out (1) if real-world requirements specifications
contain duplicated information, (2) what kind of information is duplicated,
(3) which consequences the duplication of information has on the different
software development activities, and (4) if existing clone detection
approaches can be applied in practice to identify duplication in SRS in an
automated way. In this case study we analyzed 28 requirements
specifications with a total of 8,667 pages that describe, amongst others,
business information and embedded systems. To conduct the study we applied
an adapted version of a state-of-the-art clone detection
tool~\cite{Juergens2009_clonedetective} and manually inspected the
detection results to identify the inevitable false positives. The case
study revealed that the amount of duplicated information varies
significantly between SRS; starting from documents with no duplication at
all up to documents in which more than 70\% of the information is
duplicated. On average, requirements specifications in this study were more
than 13\% larger than they would be without duplication. For one analyzed
specification, this increases the effort required for a single inspection
by 13 person days.

\minisec{Research Problem} Due to their mostly textual nature, requirements
specifications nowadays almost fully elude automated quality analysis
although they have a paramount role in the software development process.
\emph{Clone detection}, which is commonly applied to find duplication in
source code, appears to be a promising technique to identify redundancy in
requirements specifications. However, it is currently unclear how much
duplicated information requirements specifications do contain, what the
consequences of duplication are and if duplication can be detected with
existing tools in practice.

\minisec{Contributions} We extend the empirical knowledge by a case study
that investigates the amount and nature of duplicated text contained in
real-world requirements specifications. Based on this, we illustrate the
consequences that duplication has on different software development
activities. Additionally, we demonstrate that existing clone detection
approaches can be successfully adapted to identify duplication in
requirements specifications in practice. The tool used in this study is
available as open source to be applied by practitioners as well as
researchers.

\section{Terms}

We use the term \emph{requirements specification} according to IEEE Std
830-1998 \cite{ieee98} to denote \emph{a specification for a particular
software product, program, or set of programs that performs certain
functions in a specific environment}. A single specification can comprise
multiple individual documents.

A requirements specification is interpreted as a single sequence of words. In
case it comprises multiple documents, individual word lists are concatenated to
form a single list for the requirements specification. \emph{Normalization} is
a function that transforms words to remove subtle syntactic differences between
words with similar denotation. A \emph{normalized specification} is a sequence
of normalized words. A \emph{specification clone} is a (consecutive) substring
of the normalized specification with a certain minimal length, appearing at
least twice. A \emph{clone group} contains all clones of a specification that
have the same content.

For analyzing the precision of automated detection, we further distinguish
between \emph{relevant} clone groups and \emph{false positives}. Clones of a
relevant clone group must convey semantically similar information and this
information must refer to the system described. Examples of relevant clones are
duplicated use case preconditions or system interaction steps. Examples of
false positives are duplicated document headers or footers or substrings that
contain the last words of one and the first words of the subsequent sentence
without conveying meaning.

\emph{Clone coverage} denotes the part of a specification that is covered
by cloning. It approximates the probability that an arbitrarily chosen
specification sentence is cloned at least once. \emph{Number of clone
groups and clones} denotes how many different logical specification
fragments have been copied and how often they occur. \emph{Blow-up}
describes how much larger the specification is compared to a hypothetical
specification version that contains no clones.

Document fragments that convey similar information but are not similar on the
word level are not considered as clones in this paper. While such redundancy
can also be relevant for the quality assessment of requirements specifications,
it is outside the scope of this paper.

\section{Methodology}
This section describes the study that was performed to investigate cloning in
requirements specifications and its implications for software engineering in
practice.

\subsection{Study Definition}

We outline the study using the Goal-Question-Metric template as proposed
in~\cite{Wohlin.2000}. The study has the objective to characterize and
understand the phenomenon of cloning in requirements specifications. The
result is intended to help using clone detection techniques appropriately
in the quality assessment of requirements specifications. It is thus
performed from the viewpoint of requirements engineers and SRS quality
assessors. For this, the extent of cloning in requirements specifications
and its effects on software engineering activities is investigated.
Therefore, a set of specifications from industrial projects are used as
study objects. We further detail the objectives of the study using four
research questions:

\begin{question}
How much cloning do real-world requirements specifications contain?
\end{question}

\begin{question}
What kind of information is cloned in requirements specifications?
\end{question}

\begin{question}
What consequences does cloning in requirements specifications have?
\end{question}

\begin{question}
Can cloning in requirements specifications be detected accurately using
existing clone detectors?
\end{question}

\subsection{Study Design}
This section gives a high level overview of the study design. Details on its
implementation and execution are given in Sec. \ref{sec:implementation}.

The study uses content analysis of specification documents to
answer the research questions. For further explorative analyses
the content of source code is also analyzed. The content analysis is performed
using a clone detection tool as well as manually.

First, we assign requirements specifications randomly to pairs of researchers for
the further analysis. We do this to reduce any potential bias that is
introduced by the researchers. Clone detection (\cf
Sec.~\ref{sec:detection}) is performed on all documents of a specification.

Next, the researcher pairs perform clone detection tailoring for each specification. For this,
they manually inspect detected clones for false positives. Filters are
added to the detection configuration so that these false positives no longer
occur. The detection is re-run and the detected clones are analyzed. This is
repeated until no false positives are found in a random sample of the detected
clone groups. In order to answer RQ 4, precision before and after tailoring,
categories of false positives and times required for tailoring are
recorded.

The results of the tailored clone detection comprise a report with all
clones and clone metrics that are used to answer RQ 1: clone coverage,
number of clone groups and clones, and blow-up. Blow-up is measured in
relative and absolute terms. Standard values for reading and inspection
speeds from the literature are used to quantify the additional effort that
this blow-up causes. Blow-up and cloning-induced efforts are used to
answer RQ 3.

For each specification, we qualitatively analyze a random sample of
clone groups for the kind of information they contain. We start with an initial
categorization from an earlier study~\cite{streit09} and extend it, when
necessary, during categorization (formally speaking, we thus employ a mixed
theory-based and grounded theory approach \cite{Corbin.2008}). If a clone
contains information that can be assigned to more than one category, it is
assigned to all suitable categories. The resulting categorization of cloned
information in requirements specifications is used to answer RQ~2. In order to
ensure a certain level of objectiveness, inter-rater agreement is measured
for the resulting categorization.

In many software projects, SRS are no
read-only artifacts but undergo constant revisions to adapt to ever
changing requirements. Such modifications are hampered by cloning as
changes to duplicated text often need to be carried out in multiple
locations. Moreover, if the changes are unintentionally not performed to
all affected clones, inconsistencies can be introduced in SRS that later
on create additional efforts for clarification. In the worst case, they
make it to the implementation of the software system, causing
inconsistent behavior of the final product. Studies show that this occurs
in practice for inconsistent modifications to code
clones~\cite{juergens09} -- we thus expect that it can also happen in SRS.
Hence, in addition to the categories, further noteworthy issues of the clones noticed
during manual inspection are documented. Examples are inconsistencies in the
duplicated specification fragments. This information is used to for additional
answers RQ~3.

Moreover, on selected specifications, content analysis of the source code of
the implementation is performed: we investigate the code
corresponding to specification clones in order to classify whether the
specification cloning resulted in code cloning, duplicated functionality
without cloning, or was resolved through the creation of a shared abstraction.
These effects are only given qualitatively. Further quantitative analysis is
beyond the scope of this paper.

\begin{figure}[htbp]
\begin{center}
\includegraphics[width=.45\textwidth]{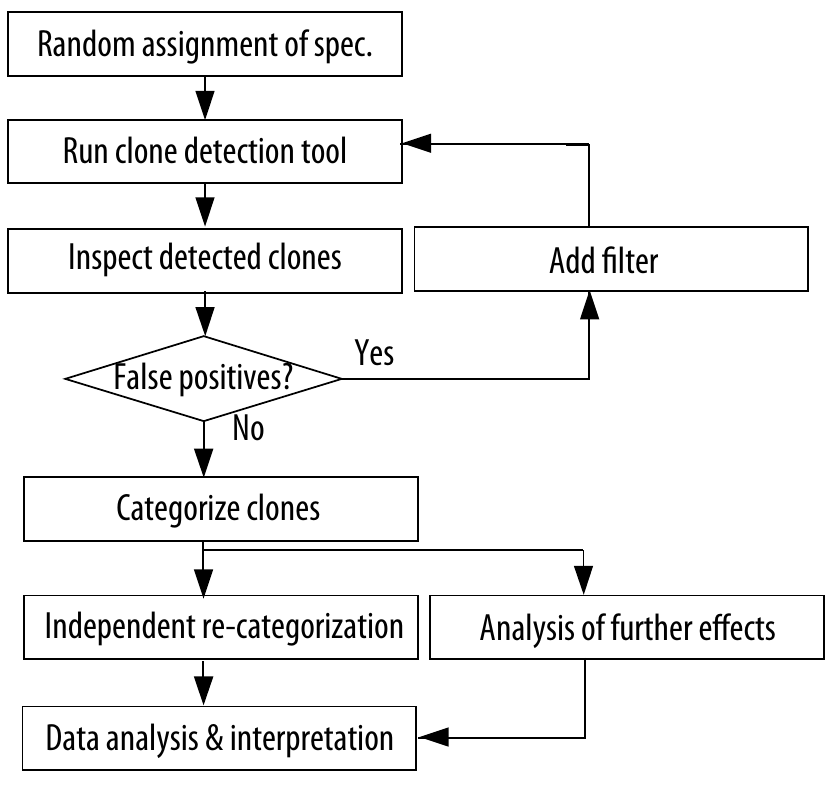}
\caption{Study design overview}
\label{fig:design_overview}
\end{center}
\end{figure}

In the final step, all collected data is analyzed and interpreted
in order to answer the research questions. An overview of the
steps of the study is given in Fig.~\ref{fig:design_overview}.

\subsection{Study Objects}

We use 28 requirements specifications as study objects from the domains of
administration, automotive, conve\-ni\-ence, finance, telecommunication, and
transportation. The specified system types comprise business information
systems, software development tools, platforms, and embedded systems. The
specifications are written in English or German. Their scope ranges
from a part to the entire set of requirements of the software systems they
describe. For non-disclosure reasons, the systems are named A to Z to AC.
An overview is given in Table \ref{tab:objects}. The specifications were
obtained from different organizations, including\footnote{Due to
non-disclosure reasons, we cannot list all 11 companies from which
specifications were obtained.}:

\begin{description}
\item[Munich Re Group] is one of the largest re-insurance
companies in the world and employs more than 47,000 people in over 50
locations. For their insurance business, they develop a variety of individual
supporting software systems.

\item[Siemens AG] is the largest engineering
  company in Europe. The specification used here was obtained from the
  business unit dealing with industrial automation.

\item[MOST Cooperation] is a partnership of car
  manufacturers and component suppliers that defined an automotive multimedia
  protocol. Key partners include Audi, BMW and Daimler.
\end{description}

The specifications mainly contain natural language text. If present, other content, such as
images or diagrams, was ignored during clone detection. Specifications N, U and Z are
Microsoft Excel documents. Since they are not organized as printable pages, no
page counts are given for them. The remaining specifications are either in
Adobe PDF or Microsoft Word format. In some cases, these specifications
are generated from requirements management tools. To the best of our
knowledge, the duplication encountered in the specifications is not
introduced during generation.

Obviously, the specifications were not sampled randomly, since we had to rely
on our relationships with our partners to obtain them. However, we selected
specifications from different companies for different types of systems in
different domains to increase generalizability of the results.

\begin{table}[htbp]
\caption{Study Objects \label{tab:objects}}
\begin{center}
{\small\sffamily
\begin{tabular}{|l|r|r||l|r|r|}
\hline
\multicolumn{1}{|l|}{\textbf{Spec}} & \multicolumn{1}{c|}{\textbf{Pages}} & \multicolumn{1}{c||}{\textbf{Words}} & \multicolumn{1}{l|}{\textbf{Spec}} & \multicolumn{1}{c|}{\textbf{Pages}} & \multicolumn{1}{c|}{\textbf{Words}}\\\hline
\hline
\rowcolor{Gray}
A & 517 & 41,482 & O & 184 & 18,750\\\hline
\rowcolor{white}
B & 1,013 & 130,968 & P & 45 & 6,977\\\hline
\rowcolor{Gray}
C & 133 & 18,447 & Q & 33 & 5,040\\\hline
\rowcolor{white}
D & 241 & 37,969 & R & 109 & 15,462\\\hline
\rowcolor{Gray}
E & 185 & 37,056 & S & 144 & 24,343\\\hline
\rowcolor{white}
F & 42 & 7,662 & T & 40 & 7,799\\\hline
\rowcolor{Gray}
G & 85 & 10,076 & U & n/a & 43,216\\\hline
\rowcolor{white}
H & 160 & 19,632 & V & 448 & 95,399\\\hline
\rowcolor{Gray}
I & 53 & 6,895 & W & 211 & 31,670\\\hline
\rowcolor{white}
J & 28 & 4,411 & X & 158 & 19,679\\\hline
\rowcolor{Gray}
K & 39 & 5,912 & Y & 235 & 49,425\\\hline
\rowcolor{white}
L & 535 & 84,959 & Z & n/a & 13,807\\\hline
\rowcolor{Gray}
M & 233 & 46,763 & AB & 3,100 & 274,489\\\hline
\rowcolor{white}
N & n/a & 103,067 & AC & 696 & 81,410\\\hline
\hline
\rowcolor{Gray}
\multicolumn{4}{|l|}{$\Sigma$} & 8,667 & 1,242,765\\\hline
\end{tabular}}
\end{center}
\vspace{-1mm}
\end{table}

\subsection{Study Implementation and Execution}
\label{sec:implementation}

This section gives detailed information on how the study design was implemented
and executed on the study objects.

For RQ 1, clone detection is performed using the tool ConQAT as described
in Sec. \ref{sec:detection}. It is also used to compute the clone measures.
Detection is performed with a minimal clone length of 20 words. This
threshold was found to provide a good balance between precision and recall
during precursory experiments in which clone detection tailoring was applied.

For RQ 2, if more than 20 clone groups are found for a specification, the
manual classification is performed on a random sample of 20 clone groups. Else,
all clone groups for a specification are inspected. During inspection, the
categorization was extended by 8 categories, 1 was changed, none were removed.
In order to improve the quality of the categorization results, categorization
is performed by a team of 2 researchers for each specification. Inter-rater
agreement is determined by calculating Cohen's Kappa. For this, from 5 randomly
sampled specifications, 5 clone groups each are independently re-categorized by
2 researchers.

For RQ 3, relative blow-up is computed as the ratio of the total number of
words to the number of redundancy-free words. Absolute blow-up is
computed as the difference of total and redundancy free number of words.
The additional effort for reading is calculated using the data from
\cite{1987_gouldj_reading_speed}, which gives an average reading speed of
220 words per minute. For the impact on inspections performed on the
requirements specifications, we refer to Gilb and Graham \cite{gilb93} that
suggest 1 hour per 600 words as inspection speed. This additional effort is
calculated for each specification as well as the mean over all.

To analyze the consequences of specification cloning on source code, we
use a convenience sample of the study objects. We cannot employ a random
sample, since for many study objects, the source code is unavailable or
traceability between SRS and source code is too poor. Of the systems with
sufficient traceability, we investigate the 5 longest and the 5 shortest
clone groups as well as the 5 clone groups with the least and the 5 with
the most instances. The requirements' IDs in these clone groups are traced
to the code and compared to clone detection results on the code level.
Code clone detection is also performed using ConQAT.

For RQ 4, precision is determined by measuring the percentage of the
relevant clones in the inspected sample. Clone detection tailoring is
performed by creating regular expressions that match the false positives.
Specification fragments that match these expressions are then excluded from
the analysis.
To keep
manual effort within feasible bounds, a maximum number of 20 randomly
chosen clone groups is inspected in each tailoring step, if more than 20
clone groups are found for a specification. Else, false positives are
removed manually and no further tailoring is performed.

\section{Results}\label{sec:results}
This section presents results ordered by research question.

\subsection{RQ 1: Amount of Cloning}

RQ 1 investigates the extent of cloning in real-world requirements
specifications. The results are shown in columns 2--4 of Table
\ref{tab:results}. Clone coverage varies widely: from specifications Q and T,
in which not a single clone of the required length is found, to specification H
containing about two-thirds of duplicated content. 6 out of the 28 analyzed
specifications (namely A, F, G, H, L, Y) have a clone coverage above 20\%. The
average specification clone coverage is 13.6\%. Specifications A, D, F, G, H,
K, L, V and Y even have more than one clone per page. No correlation between
size of the specification and cloning is found. Pearson's coefficient for clone
coverage and number of words is -0.06 and thus confirms a lack of
correlation.

\begin{table}[t]
\caption{Study Results: Cloning \label{tab:results}}
\begin{center}
{\small\sffamily
\begin{tabular}{|l||r|r|r||r|r|}
\hline

\multicolumn{1}{|l||}{\textbf{Spec}} & \multicolumn{1}{c|}{\textbf{Clone}} & \multicolumn{1}{c|}{\textbf{Clone}} & \multicolumn{1}{c||}{\textbf{clones}} & \multicolumn{1}{c|}{\textbf{blow-up}} & \multicolumn{1}{c|}{\textbf{blow-up}}\\
 & \multicolumn{1}{c|}{\textbf{cov.}} & \multicolumn{1}{c|}{\textbf{groups}} &  & \multicolumn{1}{c|}{\textbf{relative}} & \multicolumn{1}{c|}{\textbf{words}}\\\hline
\hline
\rowcolor{Gray}
A & 35.0\% & 259 & 914 & 32.6\% & 10,191\\\hline
\rowcolor{white}
B & 8.9\% & 265 & 639 & 5.3\% & 6,639\\\hline
\rowcolor{Gray}
C & 18.5\% & 37 & 88 & 11.5\% & 1,907\\\hline
\rowcolor{white}
D & 8.1\% & 105 & 479 & 6.9\% & 2,463\\\hline
\rowcolor{Gray}
E & 0.9\% & 6 & 12 & 0.4\% & 161\\\hline
\rowcolor{white}
F & 51.1\% & 50 & 162 & 60.6\% & 2,890\\\hline
\rowcolor{Gray}
G & 22.1\% & 60 & 262 & 20.4\% & 1,704\\\hline
\rowcolor{white}
H & 71.6\% & 71 & 360 & 129.6\% & 11,083\\\hline
\rowcolor{Gray}
I & 5.5\% & 7 & 15 & 3.0\% & 201\\\hline
\rowcolor{white}
J & 1.0\% & 1 & 2 & 0.5\% & 22\\\hline
\rowcolor{Gray}
K & 18.1\% & 19 & 55 & 13.4\% & 699\\\hline
\rowcolor{white}
L & 20.5\% & 303 & 794 & 14.1\% & 10,475\\\hline
\rowcolor{Gray}
M & 1.2\% & 11 & 23 & 0.6\% & 287\\\hline
\rowcolor{white}
N & 8.2\% & 159 & 373 & 5.0\% & 4,915\\\hline
\rowcolor{Gray}
O & 1.9\% & 8 & 16 & 1.0\% & 182\\\hline
\rowcolor{white}
P & 5.8\% & 5 & 10 & 3.0\% & 204\\\hline
\rowcolor{Gray}
Q & 0.0\% & 0 & 0 & 0.0\% & 0\\\hline
\rowcolor{white}
R & 0.7\% & 2 & 4 & 0.4\% & 56\\\hline
\rowcolor{Gray}
S & 1.6\% & 11 & 27 & 0.9\% & 228\\\hline
\rowcolor{white}
T & 0.0\% & 0 & 0 & 0.0\% & 0\\\hline
\rowcolor{Gray}
U & 15.5\% & 85 & 237 & 10.8\% & 4,206\\\hline
\rowcolor{white}
V & 11.2\% & 201 & 485 & 7.0\% & 6,204\\\hline
\rowcolor{Gray}
W & 2.0\% & 14 & 31 & 1.1\% & 355\\\hline
\rowcolor{white}
X & 12.4\% & 21 & 45 & 6.8\% & 1,253\\\hline
\rowcolor{Gray}
Y & 21.9\% & 181 & 553 & 18.2\% & 7,593\\\hline
\rowcolor{white}
Z & 19.6\% & 50 & 117 & 14.2\% & 1,718\\\hline
\rowcolor{Gray}
AB & 12.1\% & 635 & 1818 & 8.7\% & 21,993\\\hline
\rowcolor{white}
AC & 5.4\% & 65 & 148 & 3.2\% & 2,549\\\hline
\hline
Avg & 13.6\% &  &  & 13.5\% & \\\hline
$\Sigma$ &  & 2,631 & 7,669 &  & 100,178\\\hline

\end{tabular}}
\end{center}
\end{table}

\begin{figure}[h]
\centering
\includegraphics[width=.98\columnwidth]{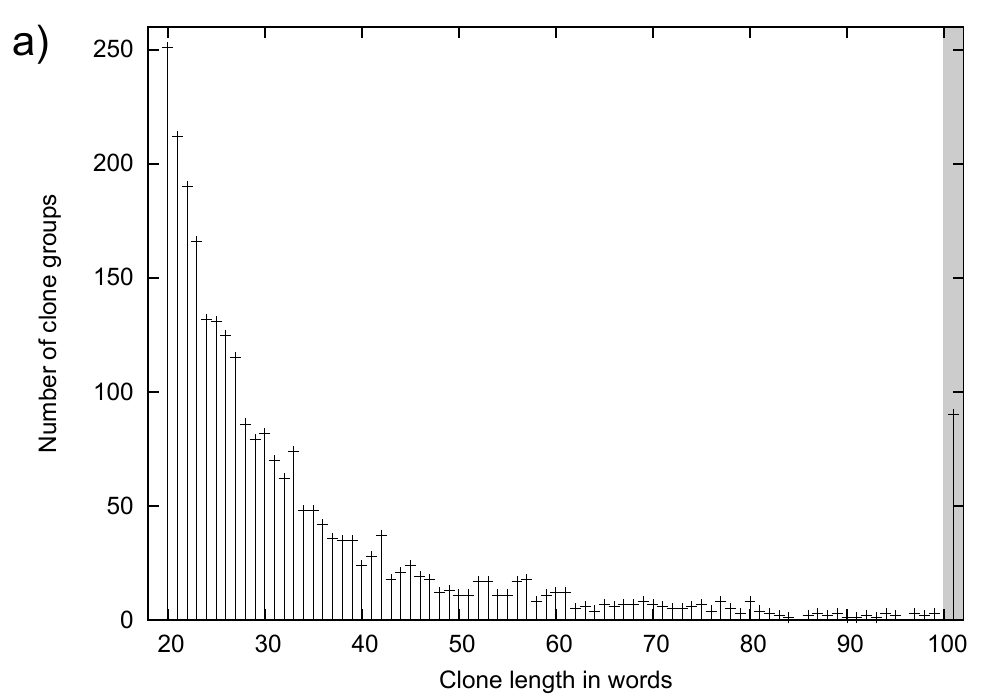}
\includegraphics[width=.98\columnwidth]{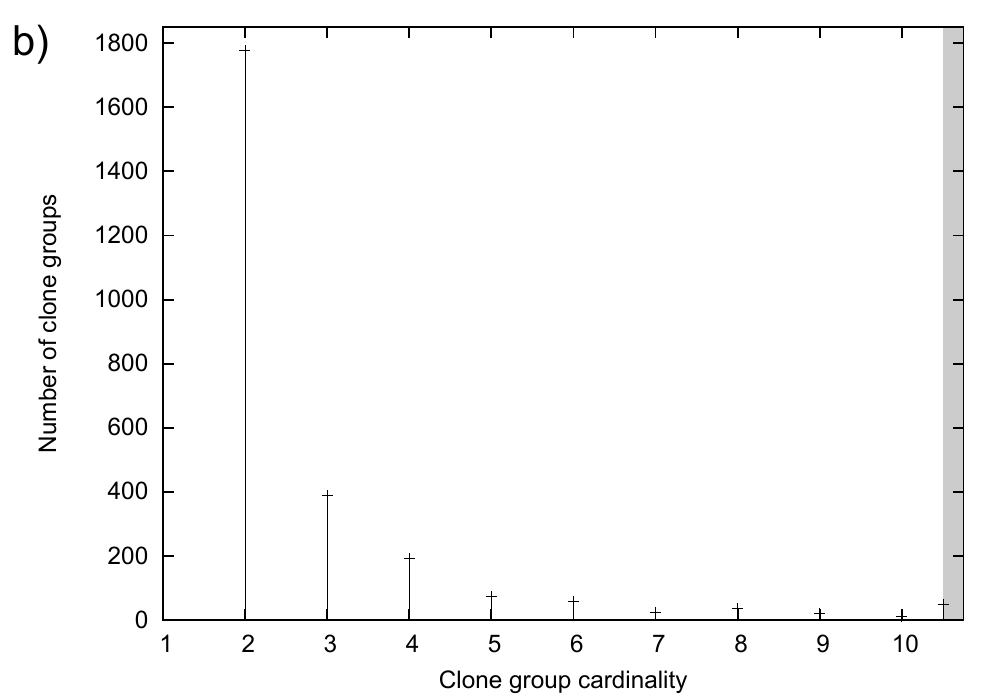}
\caption{Distribution of clone lengths and clone group cardinalities}
\label{f:clone_dists}
\end{figure}

Fig.~\ref{f:clone_dists} depicts the distribution of clone lengths
in words~(a) and of clone group cardinalities~(b), \ie the number of
times a specification fragment has been cloned\footnote{The rightmost
value in each diagram aggregates data that is outside
its range. For space reasons, the distributions are given for the
union of detected clones across specifications and not for each one
individually. The general observations are, however, consistent across
specifications.}. Short clones are more frequent than
long clones. Still, 90 clone groups have a length greater
than 100 words. The longest detected group comprises two clones of
1049 words each, describing similar input dialogs for different types of
data.

Clone pairs are more frequent than clone groups of cardinality 3 or higher.
However, 49 groups with cardinality above 10 were detected. The largest
group encountered contains 42 clones. They contain domain knowledge
about roles involved in contracts that has been duplicated 42 times.

\subsection{RQ 2: Cloned Information}
RQ 2 investigates which type of information is cloned in real-world
requirements specifications. The categories of cloned information
encountered in the study objects are:

\minisec{Detailed Use Case Steps:} Description of one or more steps in a use case that
         specifies in detail how a user interacts with the system, such as the
         steps required to create a new customer account in a system.

\minisec{Reference:} Fragment in a requirements specification that
         refers to another document or another part of the same document. Examples
         are references in a use case to other use cases or to the corresponding
         business process.

\minisec{UI:} Information that refers to the
         (graphical) user interface. The specification of which buttons are visible on
         which screen is an example for this category.

\minisec{Domain Knowledge:} Information about the application
         domain of the software. An example are details about what is part of an
         insurance contract for a software that manages insurance contracts.

\minisec{Interface Description:} Data and message definitions that describe the interface of a component, function, or system.
         An example is the definition of messages on a bus system that a component
         reads and writes.

\minisec{Pre-Condition:} A condition that has to hold before something else can happen.
         A common example are pre-conditions for the execution of a specific use case.

\minisec{Side-Condition:} Condition that describes the status that has to hold during the
         execution of something. An example is that a user has to remain logged
         in during the execution of a certain functionality.

\minisec{Configuration:} Explicit settings for configuring
         the described component or system. An
         example are timing parameters for configuring a transmission protocol.

\minisec{Feature:} Description of a piece of functionality of the system on a high level
         of abstraction.

\minisec{Technical Domain Knowledge:} Information
         about the used technology for the solution and the technical
         environment of the system, \eg used bus systems in an embedded system.

\minisec{Post-Condition:} Condition that describes what has to hold after something
         has been finished. Analogous to the pre-conditions, post-conditions are usually
         part of use cases to describe the system state after the use case execution.

\minisec{Rationale:} Justification of a requirement. An example is the explicit demand
         by a certain user group.

\vspace{2mm}

\begin{figure}[h]
\centering
\includegraphics[width=.98\linewidth]{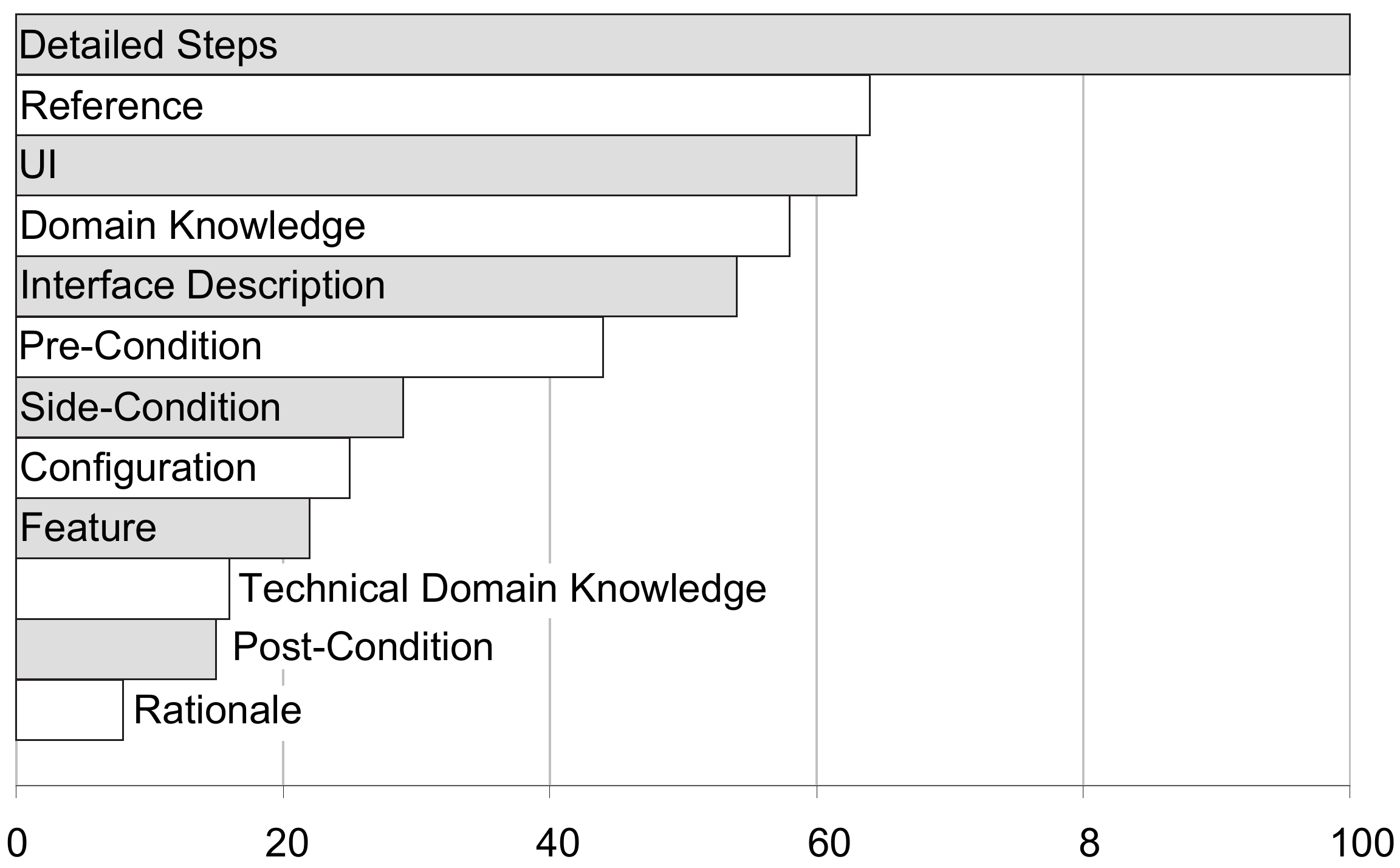}
\caption{Quantitative results for the categorization of cloned information}
\label{fig:categorization}
\end{figure}

We document the distribution of clone groups to the categories for the
sample of categorized clone groups. 404 clone groups are assigned 498
times (multiple assignments are possible). The quantitative results of the
categorization are depicted in Fig.~\ref{fig:categorization}. The highest
number of assignments are to category ``Detailed Use Case Steps'' with 100
assignments. ``Reference'' (64) and ``UI'' (63) follow. The least number of
assignments are to category ``Rationale'' (8).

The random sample for inter-rater agreement calculation consists of the
specifications L, R, U, Z, and AB. From each specification, 5 random clones
are inspected and categorized. As one specification only has 2 clone groups,
in total 22 clone groups are inspected.
We measure the inter-rater agreement
using Cohen's Kappa with a result of 0.67; this is commonly considered as
substantial agreement. Hence, the categorization is good enough to ensure that
independent raters categorize the cloned information
similarly, which implies a certain degree of completeness and suitability.

\subsection{RQ 3: Consequences of SRS Cloning}

RQ 3 investigates the consequences of SRS cloning with respect to (1)
specification reading, (2) specification modification and (3) specification
implementation, \ie activities that use SRS as an input.

\minisec{Specification Reading} Cloning in specifications obviously
increases specification size and, hence, affects all activities that
involve reading the specification documents. As
Table~\ref{tab:results_blow-up} shows, the average blow-up of the analyzed
SRS is 3,578 words which, at typical reading speed of 220 words per
minute~\cite{1987_gouldj_reading_speed}, translates to additional
$\approx$16 minutes spent on reading for each document. While this does not
appear to be a lot, one needs to consider that quality assurance techniques
like inspections usually assume a significantly lower processing rate. For
example, \cite{gilb93} considers 600 words per hour as the maximum rate for
effective inspections. Hence, the average additional time spent on
inspections of the analyzed SRS is expected to be about 6 hours. In a
typical inspection meeting with 3 participants, this amounts to 2.25 person
days. For specification AB that has a blow-up of 21,993 words, the
additional effort is expected to be greater than 13 person days
if three inspectors are applied.

\begin{table}[h]
\caption{Study Results: Consequences \label{tab:results_blow-up}}
\begin{center}
{\small\sffamily
\begin{tabular}{|l|r|r|r||l|r|r|r|}
\hline

\multicolumn{1}{|l|}{\textbf{S}} & \multicolumn{1}{c|}{\textbf{blow-up}} & \multicolumn{1}{c|}{\textbf{read.}} & \multicolumn{1}{c||}{\textbf{insp.}} & \multicolumn{1}{l|}{\textbf{S}} & \multicolumn{1}{c|}{\textbf{blow-up}} & \multicolumn{1}{c|}{\textbf{read.}} & \multicolumn{1}{c|}{\textbf{insp.}}\\
 & \multicolumn{1}{c|}{[words]} & \multicolumn{1}{c|}{[m]\footnotemark} & \multicolumn{1}{c||}{[h]\footnotemark} &  & \multicolumn{1}{c|}{[words]} & \multicolumn{1}{c|}{[m]\addtocounter{footnote}{-2}\footnotemark} & \multicolumn{1}{c|}{[h]\footnotemark}\\
\hline
\rowcolor{Gray}
A & 10,191 & 46.3 & 17.0 & O & 182 & 0.8 & 0.3\\\hline
\rowcolor{white}
B & 6,639 & 30.2 & 11.1 & P & 204 & 0.9 & 0.3\\\hline
\rowcolor{Gray}
C & 1,907 & 8.7 & 3.2 & Q & 0 & 0.0 & 0.0\\\hline
\rowcolor{white}
D & 2,463 & 11.2 & 4.1 & R & 56 & 0.3 & 0.1\\\hline
\rowcolor{Gray}
E & 161 & 0.7 & 0.3 & S & 228 & 1.0 & 0.4\\\hline
\rowcolor{white}
F & 2,890 & 13.1 & 4.8 & T & 0 & 0.0 & 0.0\\\hline
\rowcolor{Gray}
G & 1,704 & 7.7 & 2.8 & U & 4,206 & 19.1 & 7.0\\\hline
\rowcolor{white}
H & 11,083 & 50.4 & 18.5 & V & 6,204 & 28.2 & 10.3\\\hline
\rowcolor{Gray}
I & 201 & 0.9 & 0.3 & W & 355 & 1.6 & 0.6\\\hline
\rowcolor{white}
J & 22 & 0.1 & 0.0 & X & 1,253 & 5.7 & 2.1\\\hline
\rowcolor{Gray}
K & 699 & 3.2 & 1.2 & Y & 7,593 & 34.5 & 12.7\\\hline
\rowcolor{white}
L & 10,475 & 47.6 & 17.5 & Z & 1,718 & 7.8 & 2.9\\\hline
\rowcolor{Gray}
M & 287 & 1.3 & 0.5 & AB & 21,993 & 100.0 & 36.7\\\hline
\rowcolor{white}
N & 4,915 & 22.3 & 8.2 & AC & 2,549 & 11.6 & 4.2\\\hline
\hline
\rowcolor{Gray}
\multicolumn{5}{|l|}{Avg} & 3,578 & 16.3 & 6.0\\\hline

\end{tabular}}
\end{center}
\end{table}

\addtocounter{footnote}{-2}

\stepcounter{footnote}
\footnotetext{Additional reading effort in clock minutes.}

\stepcounter{footnote}
\footnotetext{Additional inspection effort in clock hours.}

\minisec{Specification Modification} To explore the extent of
inconsistencies in our specifications, we analyze the comments that were
documented during the inspection of the sampled clones for each
specification set. They refer to duplicated specification fragments that
are essentially longer than the clones detected by the tool. The full
length of the duplication is not found by the tool due to small
differences between the clones that often result from inconsistent
modification.

An example for such a potential inconsistency can be found in the publicly
available MOST specification (M). The function classes ``Sequence
Property'' and ``Sequence Method'' have exactly the same parameter lists.
They are detected as clones. The following description is also copied, but
one ends with the sentence ``Please note that in case of elements,
parameter Flags is not available.''. In the other case, this sentence is
missing. Whether these differences are really defects in the requirements
or not could only be determined by consulting the requirements engineers
of the system. This further step is out of scope of this paper.

\minisec{Specification Implementation} With respect to the entirety of the
software development process, it is important to understand which
consequences SRS cloning has on development activities that use SRS as an
input, \eg system implementation and test. For the inspected 20
specification clone groups and their corresponding source code, we found 3
different effects:

\begin{enumerate}
\item The redundancy in the requirements is not reflected in the code. It
contains shared abstractions that avoid duplication.

\item\label{item:clone} The code that implements a cloned piece of an SRS is
cloned, too. In this case, future changes to the cloned code cause additional
efforts as modifications must be reflected in all clones. Furthermore, changes
to cloned code are error-prone as inconsistencies may be introduced
accidentally \cite{juergens09}.

\item Code of the same functionality has been implemented multiple times. The
redundancy of the requirements thus does exist in the code as well but has not
been created by copy\&paste. This case exhibits similar problems as
case~\ref{item:clone} but creates additional efforts for the repeated
implementation. Moreover, this kind of redundancy is harder to detect as
existing clone detection approaches cannot find code that is functionally
similar but not the result of copy\&paste~\cite{Juergens2010_csmr}.

\end{enumerate}

\subsection{RQ 4: Detection Tailoring and Accuracy}\label{s:rq4detail}
RQ 4 investigates whether redundancy in real-world requirements specifications
can be detected with existing approaches.

Result precision values and times required for clone detection tailoring
are depicted in Table~\ref{tab:results_precision}. Tailoring times do not
include setup times and duration of the first detection run. If no clones
are detected for a specification (\ie Q and T), no precision value is
given. While for some specifications no tailoring is necessary at all, \eg
E, F, G or, S, the worst precision value without tailoring is as low as 2\%
for specification O. In this case, hundreds of clones containing only the
page footer cause the large amount of false positives. For 8
specifications (A, C, M, O, P, R, AB, and AC), precision values below 50\%
are measured before tailoring. The false positives contain information
from the following categories:

\minisec{Document meta data} comprises information about the creation process
of the document. This includes author information and document edit
histories or meeting histories typically contained at the start or end of a document.

\minisec{Indexes} do not add new information and are typically generated
automatically by text processors. Encountered examples comprise tables of
content or subject indexes.

\minisec{Page decorations} are typically automatically inserted by text
processors. Encountered examples include page headers and footers and page or
line numbers.

\minisec{Open issues} document gaps in the specification. Encountered examples
comprise ``TODO'' statements or tables with unresolved questions.

\minisec{Specification template information} contains section names common
to all individual documents that are part of a specification.

\vspace{2mm}

Some of the false positives, such as document headers or footers could possibly
be avoided by accessing requirements information in a more direct form than done
by text extraction from requirements specification documents.

Precision was increased substantially by clone detection tailoring.
Precision values for the specifications are above 85\%, average precision
is 99\%. The time required for tailoring varies between 1 and 33 minutes
across specifications. Low tailoring times occurred, when either no false
positives were encountered, or they could very easily be removed, \eg
through exclusion of page footers by adding a single simple regular
expression. On average, 10 minutes are required for tailoring.

\begin{table}[htbp]
\caption{Study Results: Tailoring \label{tab:results_precision}}
\begin{center}
{\small\sffamily
\begin{tabular}{|l|r|r|r||l|r|r|r|}
\hline

\multicolumn{1}{|l|}{\textbf{S}} & \multicolumn{1}{c|}{\textbf{Prec.}} & \multicolumn{1}{c|}{\textbf{Tail.}} & \multicolumn{1}{c||}{\textbf{Prec.}} & \multicolumn{1}{l|}{\textbf{S}} & \multicolumn{1}{c|}{\textbf{Prec.}} & \multicolumn{1}{c|}{\textbf{Tail.}} & \multicolumn{1}{c|}{\textbf{Prec.}}\\
 & \multicolumn{1}{c|}{\textbf{bef.}} & \multicolumn{1}{c|}{\textbf{min}} & \multicolumn{1}{c||}{\textbf{after}} &  & \multicolumn{1}{c|}{\textbf{bef.}} & \multicolumn{1}{c|}{\textbf{min}} & \multicolumn{1}{c|}{\textbf{after}}\\
\hline
\rowcolor{Gray}
A & 27\% & 30 & 100\% & O & 2\% & 8 & 100\%\\\hline
\rowcolor{white}
B & 58\% & 15 & 100\% & P & 48\% & 20 & 100\%\\\hline
\rowcolor{Gray}
C & 45\% & 25 & 100\% & Q & n/a & 1 & n/a\\\hline
\rowcolor{white}
D & 99\% & 5 & 99\% & R & 40\% & 4 & 100\%\\\hline
\rowcolor{Gray}
E & 100\% & 2 & 100\% & S & 100\% & 2 & 100\%\\\hline
\rowcolor{white}
F & 100\% & 4 & 100\% & T & n/a & 1 & n/a\\\hline
\rowcolor{Gray}
G & 100\% & 2 & 100\% & U & 85\% & 5 & 85\%\\\hline
\rowcolor{white}
H & 97\% & 10 & 97\% & V & 59\% & 6 & 100\%\\\hline
\rowcolor{Gray}
I & 71\% & 8 & 100\% & W & 100\% & 6 & 100\%\\\hline
\rowcolor{white}
J & 100\% & 2 & 100\% & X & 96\% & 13 & 100\%\\\hline
\rowcolor{Gray}
K & 96\% & 2 & 96\% & Y & 97\% & 7 & 100\%\\\hline
\rowcolor{white}
L & 52\% & 26 & 100\% & Z & 100\% & 1 & 100\%\\\hline
\rowcolor{Gray}
M & 44\% & 23 & 100\% & AB & 30\% & 33 & 100\%\\\hline
\rowcolor{white}
N & 100\% & 4 & 100\% & AC & 48\% & 14 & 100\%\\\hline

\end{tabular}}
\end{center}
\end{table}

\section{Discussion}

The results from the previous section imply that cloning in the sense of
copy\&paste is common in real-world requirements specifications. Here we
interpret these results and discuss their implications.

According to the results of RQ 1 the amount of cloning encountered is
significant, although the extent differs between specifications. The large
amount of detected cloning is further emphasized by the fact that our
approach only locates identical parts of the text. Other forms of
redundancy, such as specification fragments which have been copied but
slightly reworded in later editing steps or which are completely reworded,
although containing the same meaning, are not included in these numbers.
The relatively broad spectrum of findings, however, illustrates that
cloning in SRS can be successfully avoided. SRS E, for example, exhibits
almost no cloning although it is of significant size.

The results for RQ 2 illustrate that cloning is not confined to a specific
kind of information. On the contrary, we found that duplication can,
amongst others, be found in the description of use cases, the application
domain and the user interface but also in parts of documents that merely
reference other documents. Our case study only yields the absolute number
of clones assigned to a category. As we do not investigate which amount of
a SRS can be assigned to the category, we cannot deduce if cloning is more
likely to occur in one category than another. Hence, we currently assume
that clones are likely to occur in all parts of SRS.

The most obvious effect of duplication is the increased size (\cf RQ 3),
which could often be avoided by cross references or different organization
of the specifications. This affects all (manual) processing steps performed
on the specifications, such as restructuring or translating them to other
languages, and especially reading. Reading is emphasized here, as the ratio
of persons reading to those writing a specification is usually large,
even larger than in source code. The activities that involve reading
include specification reviews, system implementation, system testing, and
contract negotiations. All of them are typically performed by different
persons, which are all affected by the blow up. While the additional effort
for reading has been assumed to be linear in the presentation of the
results, one could even argue that the effort might be larger, as human
readers are not efficient with word-wise comparison. This is, however,
required to check presumably duplicated parts to find potential subtle
differences between them that could otherwise lead to errors in the final
system.

Furthermore, redundancy has effects on the quality of the specification, as
it may lead to inconsistent changes of the clones, which my induce
errors in the specification and thus often in the final system. Based on the
inconsistencies we encountered, we strongly suspect that
there is a real threat that inconsistent maintenance of duplicated SRS
introduces errors in practice. However, since we did not validate that
the inconsistencies are in fact errors, our results are not
conclusive -- future research on this topic is required.
Nevertheless, the inconsistencies probably cause
overhead during further system development due to clarification requests
from developers spotting these inconsistencies.

Moreover, our observations
show that specification cloning can lead to cloned or, even
worse, reimplemented parts of code. Often these duplications can not even
be spotted by the developers, as they only work on a part of the system,
whose sub-specification might not even contain clones when viewed in
isolation.

Redundancy is hard to identify in SRS as common quality assurance
techniques like inspections often analyze the different parts of a
specification individually and are, hence, prone to miss duplication.
The results for RQ 4 show that existing clone detection approaches can be applied to
identify cloned information in SRS in practice. However, it also shows that
a certain amount of \emph{analysis tailoring} is required to increase
detection precision. As the effort required for the tailoring steps is
below one person hour for each specification document in the case study, we
do not consider this to be an obstacle for the application of clone
detection as a means for the assessment of SRS quality in practice.

\section{Threats to Validity}

In this section, we discuss threats to the validity of the study results and
how we mitigated them.

\subsection{Internal Validity}
First of all, the results can be influenced by individual preferences or
mistakes of the researchers that performed clone detection tailoring. We
mitigated this risk by performing clone tailoring in pairs to reduce the
probability of errors and achieve better objectivity.

Precision was determined on random samples, instead of on all detected clone
groups. While this can potentially introduce inaccuracy, sampling is commonly
used to determine precision and it has been demonstrated that even small
samples can yield precise estimates
\cite{2007_Bellon_clone_detection_comparison}.

The categorization of the cloned information is subjective to some
degree. We mitigated this risk again by pairing the researchers as
well as by analyzing the inter-rater agreement as discussed in
Sec. \ref{sec:results}. All researchers were in the same room during
categorization. This way,
newly added categories were immediately available to all researchers.

The calculation of additional effort due to blow-up caused by cloning can be
inaccurate if the used data from the literature does not fit to the efforts needed
at a specific company. As the used values, however, have been confirmed
in many studies, the results should be trustworthy.

We know little about how reading speeds differ for cloned versus non-cloned
text. On the one hand, one could expect that cloned text can be read more
swiftly, since similar text has been read before. On the other hand, we noticed
in many occasions that reading cloned text can actually be a lot more time
consuming than reading non-cloned text, since the discovery and comprehension
of subtle differences is often very tedious. Lacking precise data, we treated
cloned and non-cloned text uniformly with respect to reading efforts. Further research
could help to better quantify reading efforts for cloned code.

While a lot of effort was invested into understanding detection precision,
we know less about detection recall. Firstly, if regular expressions used
during tailoring are too aggressive, detection recall can be reduced. We
used pair-tailoring and comparison of results before and after tailoring to
reduce this risk. Furthermore, we have not investigated false negatives,
\ie the amount of duplication contained in a specification and not
identified by the automated detector. The reason for this is the difficulty
of clearly defining the characteristics of such clones (having a semantic
relation but little syntactic commonality) as well as the effort required
to find them manually. The figures on detected clones are thus only a lower
bound for redundancy. While the investigation of detection recall remains
important future work, our limited knowledge about it does not affect the
validity of the detected clones and the conclusions drawn from them.

\subsection{External Validity}

The practice of requirements engineering differs strongly between different
domains, companies, and even projects. Hence, it is not clear whether the
results of this study can be generalized to all existing instances of
requirements specifications. However, we investigated 28 sets of
requirements specifications from 11 organizations with over 1.2 million
words and almost 9,000 pages. The specifications come from several
different companies, from different domains -- ranging from embedded
systems to business information systems -- and with various age and depth.
Therefore, we are confident that the results are applicable to a wide
variety of systems and domains.

\section{Detection}\label{sec:detection}

The detection of clones in the specifications, which is a prerequisite for our
case study, has been performed using a slightly extended version of the ConQAT
clone detector~\cite{Juergens2009_clonedetective}\footnote{The tool
``CloneDetective'' from the title of ~\cite{Juergens2009_clonedetective} is now part of
ConQAT.}. The necessary adaptations could be easily integrated, as the tool is
designed as an extensible analysis platform. In this section we provide an
overview on the clone detection algorithm used and especially focus on the
differences to code clone detection.

\begin{figure}
  \centering
  \includegraphics[width=.90\columnwidth]{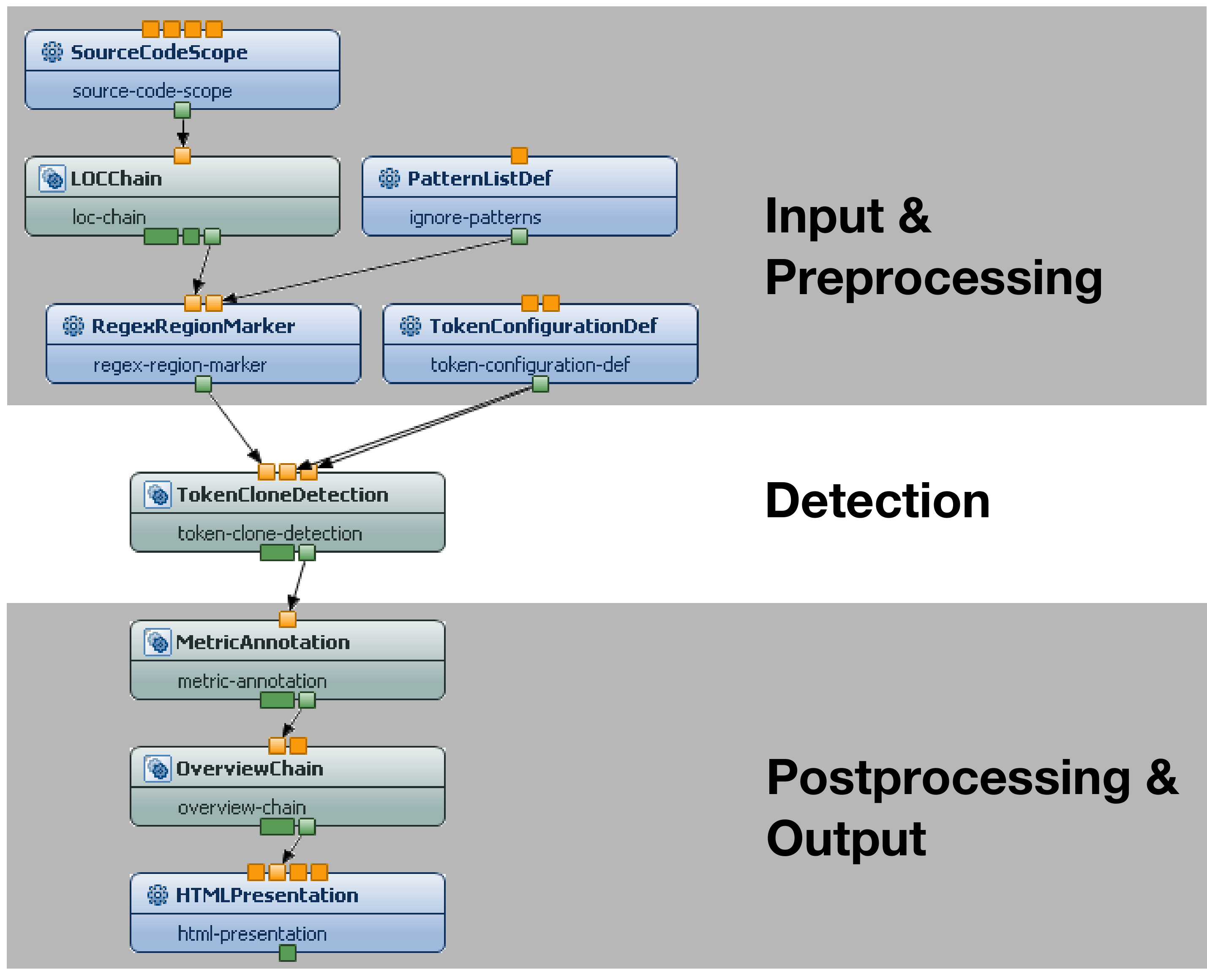}
  \caption{The clone detection pipeline}
  \label{f:pipeline}
\end{figure}

Typically, clone detection not only consists of the core algorithm,
but rather of a sequence of processing steps organized in a pipeline
fashion. This can be seen in Fig.~\ref{f:pipeline}, which depicts
the (slightly simplified) tool configuration. In the case of
ConQAT, it comes in the form of a dataflow graph. For the sake of discussion, we
group the processing steps into input \& pre-processing, detection, and
post-processing \& output, which are detailed below.

\subsection{Input and Pre-Processing}

The task of this phase is to read the documents and produce a normalized
word stream that the core algorithm then searches for identical strings of
words. Our algorithm works on plain text only, so the source documents,
which include word processor and spread sheet documents, first have to be
converted to plain text. This step is well supported by the authoring
tools, but often some clean up of the resulting text is required, \eg
discarding non-printable control characters and normalizing line breaks.
During the conversion some information is lost, such as formatting, or
diagrams and graphics. As we focus on cloning in textual descriptions, this
loss is not critical.

After reading the text contents of a specification, certain sections
of the documents are excluded. This tailoring process (\cf
Sec.~\ref{s:rq4detail}) which is specific to the specification at hand
is performed on a per document basis using regular expressions.
The resulting text is then split into single words; whitespace and
punctuation is discarded. To find clones despite minor changes, all
stop words\footnote{Stop words are used in information
  retrieval and are defined as words which are insignificant or too
  frequent to be useful in search queries. Examples are ``a'',
``and'', or ``how''. } are removed from the word list and the words are
normalized by stemming. Stemming heuristically reduces a word to its stem and
we are using the Porter stemmer algorithm \cite{porter-1980}, which is
available for various languages.  Both the list of stop words and the stemming
depend on the language of the specification.

\subsection{Detection}

In the detection phase, all substrings in the word stream that are
sufficiently long and occur at least twice are extracted. In code clone
detection there is a large number of algorithms known for this problem
\cite{2007_RoyC_Survey}; many, however, operate on data structures not
easily available for natural language texts, such as abstract syntax trees
or even on data or control flow graphs. This leaves the so called token
based approaches, which work on a sequence of tokens. The algorithm used
in our tool works by constructing a suffix tree from the token (word)
stream. Each branch of the tree which reaches at least two leaves
corresponds to a clone and is reported. However, some care has to be taken
to not report clones which are completely contained in another clone (more
details can be found in \cite{Baker1995}).

\subsection{Post-Processing and Output}

During post-processing, all clone groups which contain overlapping clones
are removed. Further filters could limit the results to clones within one
document or, vice versa, clones that span multiple documents. This kind of
filtering, however, was not used during this case study.

\begin{figure}
  \centering
  \includegraphics[width=.98\columnwidth]{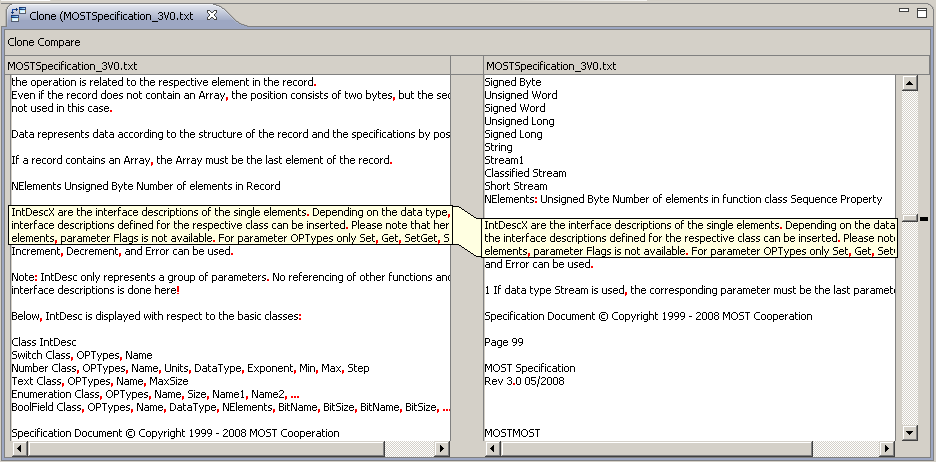}
  \caption{Screen-shot of the clone viewer application}
  \label{f:cloneviewer}
\end{figure}

The output phase calculates several metrics on the clones found,
including those reported earlier in this paper. For long-term use,
output can be rendered as an HTML report, but more crucial here is the
connection to the ConQAT clone viewer
(\cf~\cite{Juergens2009_clonedetective} and Fig.~\ref{f:cloneviewer})
to inspect the clones found in an efficient manner, which is a
prerequisite for checking for false positives and assessing the
results of the detection.

\section{Related Work}

A preliminary version of this work was published as a short paper in
\cite{streit09}. While this paper confirms the initial findings published
there, it extends and strengthens them in many important aspects:
specifications L to Z to AC have been added; the study thus now comprises
more than twice as many requirements specifications, adding up to almost
9.000 instead of the previous 2.500 pages, spanning a broader range of
domains, system types, and companies. Furthermore, a more elaborate study
design is used in this paper: clone detection tailoring is applied to
achieve better precision, which allowed lowering the minimal clone length
from 50 to 20 words and thus a more accurate analysis; a comprehensive
categorization of cloned specification fragments is given. Moreover, the
consequences of specification cloning on software engineering activities
have not been investigated in the short paper.

\minisec{Requirements quality assessment} Since accepted best practices in
requirements engineering -- like \cite{ieee98} -- recommend absence of
redundancy in requirements to ensure their modifiability, cloning is a
quality aspect of requirements specifications. In general,
\emph{structuring of requirements} and \emph{manual inspection} -- \eg
based on the criteria of  \cite{ieee98} -- are used for quality assessment
concerning redundancy. As it requires human action, it does introduce
subjectiveness and causes high expenses. While approaches exist to
mechanically analyze other quality attributes of natural language
requirements specifications -- especially ambiguity-related issues like
weak phrases, lack of imperative, or readability metrics as in, \eg,
\cite{WRH97, LLGL:2001, BGLTF:2008, KDKB:2007} -- redundancy has not been
in the focus of analysis tools. Furthermore, many of them are limited to
the English language. Structured representations of requirements allow the
calculation of more \emph{specialized metrics}~\cite{BDG04}. These can,
however, not be applied in the standard case of natural language documents.

\minisec{Commonalities detection} Algorithms for commonalities detection in
documents have been developed in several other areas.
Clustering algorithms for document retrieval, such as~\cite{WNZ01}, search
for documents about issues similar to those of a reference document.
Plagiarism detection algorithms, like~\cite{CL00, LBM04}, also address the
detection of commonalities between documents. However, while these
approaches search for commonalities between a specific document and a set
of reference documents, we also consider clones within a single document.
Still, it could be worthwhile to apply these approaches in the
context of SRS clone detection.

\minisec{Clone detection} Numerous approaches to software clone detection
have been
proposed~\cite{Koschke2008_clone_identification_removal,2007_RoyC_Survey}.
They have been applied to source
code~\cite{Koschke2008_clone_identification_removal,2007_RoyC_Survey} and
models~\cite{2008_deissenboeckf_model_clones}. Substantial investigation
of the consequences of cloning has established its negative impact on
maintenance
activities~\cite{Koschke2008_clone_identification_removal,2007_RoyC_Survey,juergens09}.
However, to our best knowledge, our work is the first to apply clone
detection approaches to requirements specifications.

\section{Conclusions and Future Work}

The IEEE standard 830-1998 requires requirements specifications to ``Not be
redundant''. Otherwise the modifiability of the document is considered
poor. Redundancy in requirements specifications is hence a factor in
assessing their quality. It has, however, not been clear
to what extent this problem exists in real-world specifications, what
consequences it has and if duplication can be detected with existing tools.
We conducted a large-scale industrial case study to address these
questions. Based on the results of our four research questions, we
formulate the following recommendations for working with SRS in practice:

\begin{itemize}

\item If nothing else is known about a requirements specification, one must
assume that the probability that an arbitrary sentence in the specification
is duplicated is greater than 10\%. Cloning is not confined to a specific
kind of information; all parts of an SRS are likely to be duplicated.
Readers of specification documents should be aware of this and pay
particular attention to subtle differences in duplicated text.

\item Due to size blow-up, cloning significantly increases the effort for
activities that involve reading of SRS, \eg inspections. Moreover, changes
to duplicated information are costly and error-prone. Cloned SRS fragments
might lead to clones in the code and/or repeated developments of the same
functionality. To prevent negative consequences, one should make SRS
authors and reviewers aware of these problems and avoid redundancy in SRS
from the beginning on. Significant amounts of cloning in SRS -- according
to our personal opinion anything above 5\% -- should be viewed as a warning
signal for potential future problems.

\item Existing clone detection approaches can be applied to detect cloning
in SRS with little effort. However, a certain amount of effort must be
spent on the tailoring of the clone detection tools to effectively reduce
rates of false positives. Provided the necessary tailoring was carried
out, we recommend to include clone detection as an integral part of SRS
inspections.

\item Sometimes duplication is employed intentionally in order to make a part of a
SRS self contained. In this case, make sure that the duplicated part is
maintained only once (\eg by using a macro mechanism in the text
processor) and that readers can recognize the duplication as such.

\end{itemize}

Coming back to the paper title, we found that automated clone detection can
indeed support the quality assessment of requirements specifications as it
is capable of identifying a well-recognized quality defect:
\emph{redundancy}.

However, a number of important research questions still remain unanswered:
how does redundancy in text documents affect reading speeds and inspection
accuracy? How can different types of redundancy be eliminated from natural
language texts? How can redundancy beyond copy\&paste be detected in
requirements specifications in a (semi-)automated fashion? Can
part-of-speech analysis techniques compensate small differences between
clones and thus increase detection recall? Apart from these research
questions, our future work is mainly aimed towards a broadening of the
scope to the whole software development life cycle. In particular, we want
to investigate more thoroughly which consequences SRS cloning has on
subsequent development activities.

\balance

\bibliographystyle{abbrv}
\bibliography{icse10}

\begin{thebibliography}{10}

\bibitem{Baker1995}
B.~S. Baker.
\newblock On finding duplication and near-duplication in large software
  systems.
\newblock In {\em WCRE'95}. IEEE, 1995.

\bibitem{2007_Bellon_clone_detection_comparison}
S.~Bellon, R.~Koschke, G.~Antoniol, J.~Krinke, and E.~Merlo.
\newblock Comparison and evaluation of clone detection tools.
\newblock {\em IEEE Trans. Softw. Eng.}, 33(9):577--591, 2007.

\bibitem{BDG04}
B.~Bern{\'a}rdez, A.~Dur{\'a}n, and M.~Genero.
\newblock Empirical evaluation and review of a metrics-based approach for use
  case verification.
\newblock {\em JRPIT}, 36(4):247--258, 2004.

\bibitem{2001_boehmb_top_10}
B.~Boehm and V.~R. Basili.
\newblock Software defect reduction top 10 list.
\newblock {\em Computer}, 34(1):135--137, 2001.

\bibitem{BGLTF:2008}
A.~Bucchiarone, S.~Gnesi, G.~Lami, G.~Trentanni, and A.~Fantechi.
\newblock {QuARS Express - A Tool Demonstration}.
\newblock In {\em ASE'08}, 2008.

\bibitem{Corbin.2008}
M.~J. Corbin and L.~A. Strauss.
\newblock {\em Basics of qualitative research: Techniques and procedures for
  developing grounded theory}.
\newblock Sage Publ., 3. edition, 2008.

\bibitem{CL00}
F.~Culwin and T.~Lancaster.
\newblock A review of electronic services for plagiarism detection in student
  submissions.
\newblock In {\em ITiCSE'00}, 2000.

\bibitem{2008_deissenboeckf_model_clones}
F.~Deissenboeck, B.~Hummel, E.~Juergens, B.~Schaetz, S.~Wagner, J.-F. Girard,
  and S.~Teuchert.
\newblock Clone detection in automotive model-based development.
\newblock In {\em ICSE'08}, 2008.

\bibitem{streit09}
C.~Domann, E.~Juergens, and J.~Streit.
\newblock The curse of copy\&paste -- {C}loning in requirements specifications.
\newblock In {\em ESEM'09}, 2009.

\bibitem{LLGL:2001}
F.~Fabbrini, M.~Fusani, S.~Gnesi, and G.~Lami.
\newblock {An Automatic Quality Evaluation for Natural Language Requirements}.
\newblock In {\em REFSQ'01}, 2001.

\bibitem{gilb93}
T.~Gilb and D.~Graham.
\newblock {\em Software Inspection}.
\newblock Addison-Wesley, 1993.

\bibitem{1981_glassr_re_defects}
R.~L. Glass.
\newblock Persistent software errors.
\newblock {\em IEEE Trans. Softw. Eng.}, 7(2):162--168, 1981.

\bibitem{1987_gouldj_reading_speed}
J.~D. Gould, L.~Alfaro, R.~Finn, B.~Haupt, and A.~Minuto.
\newblock Why reading was slower from {CRT} displays than from paper.
\newblock {\em SIGCHI Bull.}, 17(SI):7--11, 1987.

\bibitem{ieee98}
{IEEE}.
\newblock Recommended practice for software requirements specifications.
\newblock Standard 830-1998, {IEEE}, 1998.

\bibitem{KDKB:2007}
L.~K. Ishrar~Hussain, Olga~Ormandjieva.
\newblock Automatic quality assessment of {SRS} text by means of a
  decision-tree-based text classifier.
\newblock In {\em QSIC'07}, 2007.

\bibitem{Juergens2009_clonedetective}
E.~Juergens, F.~Deissenboeck, and B.~Hummel.
\newblock {CloneDetective} -- {A} workbench for clone detection research.
\newblock In {\em ICSE'09}, 2009.

\bibitem{Juergens2010_csmr}
E.~Juergens, F.~Deissenboeck, and B.~Hummel.
\newblock Code similarities beyond copy \& paste.
\newblock In {\em CSMR'10}, 2010.

\bibitem{juergens09}
E.~Juergens, F.~Deissenboeck, B.~Hummel, and S.~Wagner.
\newblock Do code clones matter?
\newblock In {\em ICSE'09}, 2009.

\bibitem{Koschke2008_clone_identification_removal}
R.~Koschke.
\newblock Identifying and removing software clones.
\newblock In T.~Mens and S.~Demeyer, editors, {\em Software Evolution}.
  Springer, 2008.

\bibitem{LBM04}
C.~Lyon, R.~Barrett, and J.~Malcolm.
\newblock A theoretical basis to the automated detection of copying between
  texts, and its practical implementation in the ferret plagiarism and
  collusion detector.
\newblock In {\em Plagiarism: Prevention, Practice and Policies Conference},
  2004.

\bibitem{porter-1980}
M.~F. Porter.
\newblock An algorithm for suffix stripping.
\newblock {\em Program}, 14(3):130--137, 1980.

\bibitem{2007_RoyC_Survey}
C.~K. Roy and J.~R. Cordy.
\newblock A survey on software clone detection research.
\newblock Technical Report 2007-541, School of Computing Queen's University,
  2007.

\bibitem{WW02}
M.~Weber and J.~Weisbrod.
\newblock Requirements engineering in automotive development -- {E}xperiences
  and challenges.
\newblock In {\em RE'02}, 2002.

\bibitem{WNZ01}
J.-R. Wen, J.-Y. Nie, and H.-J. Zhang.
\newblock Clustering user queries of a search engine.
\newblock In {\em WWW'01}, 2001.

\bibitem{WRH97}
W.~M. Wilson, L.~H. Rosenberg, and L.~E. Hyatt.
\newblock Automated analysis of requirement specifications.
\newblock In {\em ICSE'97}, 1997.

\bibitem{Wohlin.2000}
C.~Wohlin, P.~Runeson, and M.~H\"{o}st.
\newblock {\em Experimentation in software engineering: An introduction}.
\newblock Kluwer Academic, Boston, Mass., 2000.

\end{thebibliography}

\end{document}